\documentclass[12pt]{article}
\usepackage{graphicx}
\usepackage{amsmath}
\usepackage{amsfonts}
\usepackage{amssymb}
\newtheorem{theorem}{Theorem}[section]

\newtheorem{lemma}[theorem]{Lemma}

\newtheorem{corollary}[theorem]{Corollary}
\newtheorem{remark}[theorem]{Remark}
\newenvironment{proof}[1][Proof]{\textsc{#1.} }{\ \rule{0.5em}{0.5em}}
\numberwithin{equation}{section}

\begin{document}

\title{The Einstein--scalar field constraints on asymptotically Euclidean manifolds}
\author{Yvonne Choquet-Bruhat, James Isenberg
\and and Daniel Pollack}

\date{June 20, 2005}
\maketitle
\begin{abstract}
We use the conformal method to obtain solutions of the Einstein--scalar field
gravitational constraint equations. Handling scalar fields is a bit more
challenging than handling matter fields such as fluids, Maxwell fields or 
Yang--Mills fields, because the scalar field introduces three extra terms 
into the
Lichnerowicz equation, rather than just one. Our proofs are constructive and
allow for arbitrary dimension ($>2)$ as well as low regularity initial data.
\end{abstract}

\textit{Dedicated to the memory of S. S. Chern, with admiration for his
mathematical discoveries and his character.}

\section{Introduction.}

To explain recent observations of far away stars and galaxies, as well as the
possible origin of matter elements, it has become more and more relevant in
Einsteinian cosmology to admit the existence of a scalar field with a
potential which remains to be estimated. On the other hand various
considerations, in particular the search for the unification of all the
fundamental fields, including gravitation, leads to the belief that the
universe has extra dimensions, beyond the usual three space and one time.
These extra dimensions would be spacelike, and their extent so small that we
don't perceive them at the usual scales of our experiments.

The relevant equations for cosmology would then be the Einstein equations on
an $n+1$ dimensional manifold $V$, with source a scalar field $\psi$ of
potential $V(\psi)$. These equations are, for a metric $g$ on $V$ of
Lorentzian signature\footnote{{\footnotesize We choose the signature to be
}$-++....+.$},
\begin{equation}
\text{\textit{Ei}}nstein(g)\equiv Ricci(g)-\frac{1}{2}R(g)=T;
\end{equation}
that is, in a local frame
\begin{equation}
S_{\alpha\beta}\equiv R_{\alpha\beta}-\frac{1}{2}g_{\alpha\beta}%
R=T_{\alpha\beta}%
\end{equation}
where $T$ is the stress energy tensor of a scalar field $\psi$ with potential
$V(\psi),$ i.e.,
\begin{equation}
T_{\alpha\beta}\equiv\partial_{\alpha}\psi\partial_{\beta}\psi-\frac{1}%
{2}g_{\alpha\beta}\partial_{\lambda}\psi\partial^{\lambda}\psi-g_{\alpha\beta
}V(\psi).
\end{equation}
The Einstein tensor satisfies the contracted Bianchi identities
\begin{equation}
\nabla_{\alpha}S^{\alpha\beta}\equiv0.
\end{equation}
The field $\psi$ is supposed to satisfy the semi linear wave equation
\begin{equation}
\nabla^{\alpha}\partial_{\alpha}\psi-V^{\prime}(\psi)=0,\text{ \ \ \ }%
V^{\prime}(\psi):=\frac{dV(\psi)}{d\psi}.
\end{equation}
The tensor $T$ is then divergence free
\begin{equation}
\nabla_{\alpha}T^{\alpha\beta}=0.
\end{equation}
As a consequence of condition 1.6, equations 1.2 are compatible.

The Cauchy problem for the Einstein equations, determination of an Einsteinian
spacetime from initial data on a spacelike $n$ dimensional manifold, is a
geometric analysis problem. Its solution does not exist for arbitrary initial
data, and is not unique from the point of view of analysis due to the
invariance of the equations under diffeomorphisms. The geometric initial data
are a triple $(M,\bar{g},K)$ with $M$ an $n$ dimensional maifold, which we
suppose to be smooth, $\bar{g}$ a Riemannian metric on $M$, and $K$ a
symmetric 2 - tensor on $M$. The Cauchy data for the scalar field are two
functions $\bar{\psi}$ and $\bar{\pi}$. An $n+1$ dimensional spacetime $(V,g)$
together with a scalar function $\psi$ on $V$ is called an Einstein scalar
development of these initial data if $M$ can be embedded in $V,$ so that $g$
induces on $M$ the metric $\bar{g}$ and $K$ can be identified with the
extrinsic curvature of $M$ as submanifold of $(V,g),$ while $\bar{\psi}$ is
the value of $\psi$ on $M,$ and $\bar{\pi}$ is the value on $M$ of the
derivative of $\psi$ in the direction of the unit normal to $M$ in $(V,g).$

In sections 1 to 5 of this article we use the conformal method to obtain an
elliptic system for the constraints satisfied by the initial data of an
Einstein - scalar field system. In the following sections we prove some
existence and uniqueness theorems for their solution in the case where
$(M,\bar{g})$ is asymptotically euclidean, under low regularity hypothesis.

The cases of compact $M$ and of $(M,\bar{g})$ asymptotically hyperbolic will
be treated elsewhere.

\section{Constraints for the Einstein - scalar field equations.}

The constraint equations are a consequence of the Gauss Codazzi identities
satisfied by the Ricci tensor of any pseudo riemannian manifold.

It is convenient to suppose that $V=M\times R$ and to choose on $V$ a moving
frame $\theta^{\alpha},$ $\alpha=0,1,...n,$called a Cauchy adapted frame as
long as $\theta^{0}$ annihilates vectors tangent to submanifolds
$M\times\{t\}.$ The space time metric is then decomposed as follows
\begin{equation}
g\equiv-N^{2}(\theta^{0})^{2}+g_{ij}\theta^{i}\theta^{j}\text{ \ \ \ with
\ \ \ }\theta^{0}\equiv dt,\text{ \ }\theta^{i}\equiv dx^{i}+\beta
^{i}dt,\text{ }i=1,...,n.
\end{equation}
The function $N$ is called the lapse and the time dependent spatial vector
$\beta$ the shift of the chosen representation of the spacetime metric. In
this frame the unit normal $n$ to a submanifold $M\times\{t\}$ has components
\begin{equation}
n^{0}=N^{-1},\text{ \ \ \ \ \ \ }n_{0}=-N,\text{ \ \ \ \ }n^{i}=n_{i}=0.
\end{equation}
The derivative of the function $\psi$ in the direction of $n\,$is
\begin{equation}
\pi\equiv N^{-1}\partial_{0}\psi,
\end{equation}
with $\partial_{0}$ the Pfaff derivative with respect to the 1-form
$\theta^{0}$ in the frame $\theta^{\alpha},$ i.e.
\begin{equation}
\partial_{0}\equiv\frac{\partial}{\partial t}-\beta^{i}\partial_{i},\text{
\ \ }\partial_{i}\equiv\frac{\partial}{\partial x^{i}}.
\end{equation}

In a Cauchy adapted frame the constraints read as the following equations,
where we overbar values induced on $M$ by spacetime quantities, and we set
\begin{equation}
\tau\equiv tr_{\bar{g}}K:=\bar{g}^{ij}K_{ij},\text{ \
$\vert$%
}K|_{\bar{g}}^{2}:=\bar{g}^{ih}\bar{g}^{jk}K_{ij}K_{hk}.
\end{equation}

\begin{itemize}
\item  Hamiltonian constraint.
\begin{equation}
R(\bar{g})-|K|_{\bar{g}}^{2}+\tau^{2}=2\rho\equiv2\bar{N}^{-2}\bar{T}_{00}.
\end{equation}

\item  Momentum constraint:
\begin{equation}
\bar{\nabla}_{i}K^{ij}-\bar{g}^{ij}\partial_{i}\tau=J^{j}\equiv-\bar{N}%
^{-1}\bar{T}_{0}^{j}.
\end{equation}
\end{itemize}

In the case under study, where a source is a scalar field $\psi$ we find that
\[
2\bar{N}^{-2}\bar{T}_{00}=\bar{\pi}^{2}+|D\bar{\psi}|_{\bar{g}}^{2}%
+2V(\bar{\psi})
\]
and
\[
-\bar{N}^{-1}\bar{T}_{0}^{i}=-\bar{\pi}\bar{g}^{ij}\partial_{j}\bar{\psi}.
\]

\section{Conformal formulation.}

\subsection{Hamiltonian constraint.}

In order to turn the Hamiltonian constraint into a semilinear elliptic
equation to be solved for a scalar function, one considers the metric $\bar
{g}$ as determined only up to a conformal factor. One sets for $n>2$
\begin{equation}
\bar{g}=\varphi^{\frac{4}{n-2}}\gamma,\text{ \ \ i.e. \ \ }\bar{g}%
_{ij}=\varphi^{\frac{4}{n-2}}\gamma_{ij}%
\end{equation}
with $\gamma$ a given Riemannian metric on $M$. This particular conformal
weight turns into a linear operator the differential operator on $\varphi$
appearing in the parenthesis of 3.2\ below.

The scalar curvatures $R(\bar{g})$ and $R(\gamma)$ of the conformal metrics
$\bar{g}$ and $\gamma$ are linked by the formula, where $\Delta_{\gamma}$ is
the Laplace operator in the metric $\gamma,$
\begin{equation}
R(\bar{g})\equiv\varphi^{-\frac{n+2}{n-2}}(\varphi R(\gamma)-{\frac
{4(n-1)}{n-2}}\Delta_{\gamma}\varphi).
\end{equation}
The Hamiltonian constraint becomes, when $\gamma$ and $K$ are known, a semi
linear elliptic equation for $\varphi$ with a non linearity of a fairly simple
type:
\begin{equation}
\Delta_{\gamma}\varphi-k_{n}R(\gamma)\varphi+k_{n}(|K|_{\bar{g}}^{2}-\tau
^{2}+2\rho)\varphi^{\frac{n+2}{n-2}}=0
\end{equation}
with
\begin{equation}
k_{n}={\frac{n-2}{4(n-1)}}.
\end{equation}

\subsection{Momentum constraint.}

We can express the momentum constraint in terms of $\gamma,$ $K,$ $\rho,J$ and
$\varphi$ by using the relations between the connections of two conformally
related metrics.

\begin{lemma}
On an $n$ dimensional manifold, if $\bar{g}=\varphi^{\frac{4}{n-2}}\gamma,$
and if the covariant derivatives in $\bar{g}$ and $\gamma$ are written
respectively as\ $\bar{\nabla}$ and $D,$ then the divergences in the metric
$\bar{g}$ and $\gamma$ of an arbitrary contravariant 2- tensor $P^{ij}$ are
linked by the identity
\begin{equation}
\bar{\nabla}_{i}P^{ij}\equiv\varphi^{-\frac{2(n+2)}{n-2}}D_{i}\{\varphi
^{\frac{2(n+2)}{n-2}}P^{ij}\}-\frac{2}{n-2}\varphi^{-1}\gamma^{ij}\partial
_{i}\varphi tr_{\gamma}P.
\end{equation}
\end{lemma}

\begin{proof}
The proof follows from a simple computation using the identity which links the
coefficients of the connections $\bar{\Gamma}$ of $\bar{g}$ and $C$ of
$\gamma$:
\begin{equation}
\bar{\Gamma}_{jh}^{i}=C_{jh}^{i}+\frac{2}{n-2}\varphi^{-1}\{\delta_{j}%
^{i}\partial_{h}\varphi+\delta_{h}^{i}\partial_{j}\varphi-\gamma^{ik}%
\gamma_{jh}\partial_{k}\varphi\}.
\end{equation}
\end{proof}

One sees from the identity 3.5 that it is convenient to split the unknown $K$
into a weighted traceless part and its trace, namely we set
\begin{equation}
K^{ij}=\varphi^{-\frac{2(n+2)}{n-2}}\tilde{K}^{ij}+\frac{1}{n}\bar{g}^{ij}%
\tau.
\end{equation}
Here $\tilde{K}^{ij}$ is a symmetric traceless two tensor, in the sense that
\begin{equation}
tr\tilde{K}\equiv\bar{g}_{ij}\tilde{K}^{ij}=\gamma_{ij}\tilde{K}^{ij}=0,
\end{equation}
while $\tau$ is the trace.

The momentum constraint 2.7\ then becomes
\begin{equation}
D_{i}\tilde{K}^{ij}=\frac{n-1}{n}\varphi^{\frac{2n}{n-2}}\gamma^{ij}%
\partial_{i}\tau+\varphi^{\frac{2(n+2)}{n-2}}J^{j}.
\end{equation}
\smallskip

It follows from an elementary computation that
\begin{equation}
|K|_{\bar{g}}^{2}\equiv\bar{g}_{ih}\bar{g}_{jk}K^{ij}K^{hk}=\varphi
^{\frac{-3n+2}{n-2}}|\tilde{K}|_{\gamma}^{2}+\frac{1}{n}\tau,\text{ \ with
\ \ }|\tilde{K}|_{\gamma}^{2}\equiv\gamma_{ih}\gamma_{jk}\tilde{K}^{ij}%
\tilde{K}^{hk}.
\end{equation}
The Hamiltonian constraint therefore reads
\begin{align}
&  \Delta_{\gamma}\varphi-k_{n}R(\gamma)\varphi+k_{n}\varphi^{\frac
{-3n+2}{n-2}}|\tilde{K}|_{\gamma}^{2}-{\frac{n-2}{4n}}\varphi^{\frac{n+2}%
{n-2}}\tau^{2}\\
&  =-\frac{n-2}{2(n-1)}\rho\varphi^{\frac{n+2}{n-2}}\nonumber
\end{align}
If $\gamma,\tilde{K},\tau$ and $\rho$ are specified, this is a semilinear
elliptic equation for $\varphi$ when $\tilde{K}$ is known, called a
Lichnerowicz equation.\textbf{\footnote{{\footnotesize This equation was
derived by Lichnerowicz 1944 for n=3, [Li]. In 1972 York [Yo72] introduced the
scaling of the sources, and in 1987 Choquet-Bruhat extended the analysis to
general n. In view of this history we refer to 3.11\ as the Lichnerowicz
equation.}}}.

\section{Scaling of $\bar{\pi}$.}

We denote by an overbar the values induced on $M$ by spacetime quantities. The
initial data of the scalar field $\psi$ is the value $\bar{\psi}$ induced by
$\psi$ on $M.$ It is independent on the choice of the conformal metric
$\gamma,$ but there is an ambiguity for the data of the initial data for
$\pi,$ because $\pi$ depends on the lapse $N:$ it holds that $\bar{\pi}%
=\bar{N}^{-1}\overline{\partial_{0}\psi}.$ We associate to the unphysical
metric $\gamma$ an unphysical lapse $\tilde{N},$ such that $\bar{N}$ and
$\tilde{N}$ have the same associated densities respectively for $\bar{g}$ and
$\gamma,$ that is:
\begin{equation}
\bar{N}(Det\bar{g})^{-\frac{1}{2}}=\tilde{N}(Det\gamma)^{-\frac{1}{2}}%
\end{equation}
i.e.
\begin{equation}
\bar{N}=\varphi^{2n/(n-2)}\tilde{N}.
\end{equation}
and we suppose that the given initial data is
\[
\tilde{\pi}=\tilde{N}^{-1}\overline{\partial_{0}\psi}=\varphi^{2n/(n-2)}%
\bar{N}^{-1}\overline{\partial_{0}\psi}=\varphi^{2n/(n-2)}\bar{\pi}.
\]

\subsection{Hamiltonian constraint.}

The energy density on $M$ of a scalar field $\psi$ with potential $V(\psi), $
for an observer at rest in the physical metric $\bar{g}$ reads as follows in
terms of the given data:
\begin{equation}
\rho=\frac{1}{2}(\varphi^{\frac{-4n}{n-2}}|\tilde{\pi}|^{2}+\varphi^{\frac
{-4}{n-2}}\gamma^{ij}\partial_{i}\bar{\psi}\partial_{j}\bar{\psi})+V(\bar
{\psi}).
\end{equation}
We see that the term $|\tilde{\pi}|^{2}$ adds in the Hamiltonian constraint
to
$\vert$%
$\tilde{K}|_{\gamma}^{2}$, while the term $V(\bar{\psi}) $ remains unscaled by
a power of $\varphi$. The $\partial\psi$ term adds a positive contribution to
the $\varphi$ term, adding to $-R(\gamma)$. The Hamiltonian constraint now
reads
\begin{equation}
\mathcal{H}\equiv\Delta_{\gamma}\varphi-f(\varphi)=0,\text{ \ }%
\end{equation}
with
\[
f(\varphi)\equiv r\varphi-a\varphi^{-\frac{3n-2}{n-2}}+b\varphi^{\frac
{n+2}{n-2}},
\]
where we have again set $k_{n}=\frac{n-2}{4(n-1)}$ and where
\begin{equation}
r\equiv k_{n}[R(\gamma)-|D\bar{\psi}|_{\gamma}^{2}],\text{\ \ }a\equiv
k_{n}(|\tilde{K}|_{\gamma}^{2}+|\tilde{\pi}|^{2}),\text{ \ }b\equiv\frac
{n-2}{4n}\tau^{2}-\frac{n-2}{(n-1)}V(\bar{\psi}).
\end{equation}
We observe that $a\geq0,$ while $b\leq0$ if $V(\bar{\psi})\geq0$ and $\tau=0$
(maximal slicing).

We call the equation 4.4 the conformally formulated Hamiltonian constraint, or
the Lichnerowicz equation for the Einstein - scalar field theory..

\subsection{Momentum constraint$_{{}}$}

The expression of the scalar field momentum density in terms of the new data
is:
\begin{equation}
J^{i}=-\bar{g}^{ij}(\partial_{j}\bar{\psi})\bar{\pi}=-\varphi^{-\frac
{2(n+2)}{n-2}}\gamma^{ij}(\partial_{j}\bar{\psi})\tilde{\pi}.
\end{equation}
The momentum constraint now reads
\begin{equation}
\mathcal{M}^{j}\equiv D_{i}\tilde{K}^{ij}-F^{j}=0
\end{equation}
with
\[
F^{j}\equiv\frac{n-1}{n}\varphi^{2n/(n-2)}\gamma^{ij}\partial_{i}\tau
-\gamma^{ij}\partial_{j}\bar{\psi}\tilde{\pi}.
\]
We call this equation the conformally formulated momentum constraint.

We have proved the following theorem.

\begin{theorem}
The conformally formulated momentum constraint of the Einstein - scalar field
system 4.7 is a linear system for $\tilde{K}$ when $\gamma,$ $\tau,\bar{\psi}$
and $\tilde{\pi}$ are given and the function $\varphi$ is known. It does not
contain $\varphi$ if $\tau$ is a constant.
\end{theorem}

\subsection{Conformal covariance of the constaint equations.}

It follows from the analysis above that that if ($\varphi,\tilde{K}$)
satisfies the conformally formulated constraints (3.9, 4.6, 3.11), for a
specified choice of the free data ($\gamma,\tau,\tilde{\psi},\tilde{\pi}),$
then
\begin{equation}
\bar{g}_{ij}=\varphi^{\frac{4}{n-2}}\gamma_{ij},\text{ \ }K^{ij}%
=\varphi^{-2(n+2)/(n-2)}\tilde{K}^{ij}+\frac{1}{n}\bar{g}^{ij}\tau,\text{
\ }\bar{\psi},\text{ }\bar{\pi}=\varphi^{\frac{2n}{n-2}}\tilde{\pi}.
\end{equation}
is a solution of the original Einstein - scalar field constraints.

The following conformal covariance result is an immediate corollary:

\begin{theorem}
Let ($\varphi,\tilde{K})$ be a solution of the conformally formulated
constraints in the metric $\gamma$ with data $\tau,$ $\bar{\psi}$ and
$\tilde{\pi}.$ Then ($\varphi^{\prime}=\theta^{-1}\varphi,$ $\tilde{K}%
^{\prime}\equiv\theta^{-2(n+2)/(n-2)}\tilde{K})$ is a solution of the
conformally formulated constraints in the metric $\gamma^{\prime}%
=\theta^{\frac{4\leq}{n-2}}\gamma$ with data $\tau^{\prime}=\tau,$ $\bar{\psi
}=\bar{\psi}^{\prime},$ $\tilde{\pi}^{\prime}=\theta^{\frac{-2n}{n-2}}%
\tilde{\pi}.$
\end{theorem}

\section{Solution of the conformal momentum constraint.}

The general solution of a non homogeneous linear system is obtained by adding
a particular solution to the general solution of the associated linear
homogeneous system, which, in the case of 3.9, is the following:
\begin{equation}
D_{j}\tilde{K}^{ij}=0,\text{ \ \ \ }\gamma_{ij}\tilde{K}^{ij}=0.
\end{equation}
\smallskip Symmetric 2- tensors satisfying 5.1 are called TT tensors
(transverse, traceless). As a consequence of lemma 3.1 the space of TT tensors
is the same for two conformal metrics.

We may obtain both the particular solution to 3.9 and the general solution to
5.1 by essentially the same ansatz. One can look for the particular solution
of 4.7 as the conformal Lie derivative of a vector field $Z,$ an element of
the formal $L^{2}$ dual of the space of TT tensors defined by
\begin{equation}
(\mathcal{L}_{\gamma,conf}Z)_{ij}:=D_{i}Z_{j}+D_{j}Z_{i}-\frac{2}{n}%
\gamma_{ij}D_{h}Z^{h}.
\end{equation}
We look for $\tilde{K}_{TT}$ as the sum of the conformal Lie derivative of a
vector $Y$ and an arbitary traceless symmetric 2 tensor $U.$ Then, setting
$X:=Z+Y$ it holds that:
\begin{equation}
\tilde{K}^{ij}=(\mathcal{L}_{\gamma,conf}X)^{ij}-U^{ij},
\end{equation}
with $X$ a vector field solution of the linear system
\begin{equation}
(\Delta_{\gamma,conf}X)^{j}:=D_{i}(\mathcal{L}_{\gamma,conf}Z)^{ij}%
=D_{i}U^{ij}+\frac{n-1}{n}\varphi^{2n/(n-2)}\gamma^{ij}\partial_{i}\tau
-\gamma^{ij}\partial_{j}\bar{\psi}\tilde{\pi}.
\end{equation}
The arbitrary data in the traceless tensor $\tilde{K}$ is the symmetric
traceless tensor $U.$

It has been noted by York [Yo99] that, though the formulation 4.4, 4.7 is
invariant in the sense of Theorem 4.2, the splitting of the solution
$\tilde{K}$ into a given traceless tensor $U$ and the conformal Lie derivative
of an unknown vector $X$ cannot be made conformally invariant. To try to
obtain $\tilde{K}^{\prime}\equiv\theta^{-2(n+2)/(n-2)}\tilde{K}$ given
$\gamma^{\prime}$=$\theta^{\frac{4}{n-2}}\gamma,$ we can impose the relation
between the given traceless tensors $U$ and $U^{\prime}:$%
\begin{equation}
U^{\prime ij}\equiv\theta^{-2(n+2)/(n-2)}U^{ij};
\end{equation}
however for an arbitrary vector $X$ one has
\[
(\mathcal{L}_{\gamma^{\prime},conf}X)^{ij}\equiv\theta^{-4/(n-2)}%
(\mathcal{L}_{\gamma,conf}X)^{ij}.\text{ }%
\]
There is no scaling of $X$ by a power of $\varphi$ that leads to a vector
$X^{\prime}$ and results in the desired scaling of its conformal Lie
derivative. York has proposed to remedy this defect by what he called ''the
conformal thin sandwich formulation'' .\ Inspired by his work, and by the
expression $K^{ij}\equiv N^{-1}\bar{\partial}_{0}g^{ij},$ we replace the
search for a particular solution as a conformal Lie derivative by the
following. For $\tilde{N}$ is a given scalar we define:
\begin{equation}
\mathcal{\tilde{L}}_{\gamma,conf}X:=\tilde{N}^{-1}\mathcal{L}_{\gamma,conf}X,
\end{equation}%
\begin{equation}
(\tilde{\Delta}_{\gamma,conf}X)^{j}:=D_{i}(\mathcal{\tilde{L}}_{\gamma
,conf}X)^{ij}.
\end{equation}
The mathematical properties of $\Delta_{\gamma,conf}$ and $\tilde{\Delta
}_{\gamma,conf}$ are essentially the same.

We choose $X$ to be a solution of the equation
\begin{equation}
(\tilde{\Delta}_{\gamma,conf}X)^{j}=D_{i}U^{ij}+\frac{n-1}{n}\varphi
^{2n/(n-2)}\gamma^{ij}\partial_{i}\tau-\gamma^{ij}\partial_{j}\bar{\psi}%
\tilde{\pi}.
\end{equation}
(instead of 5.4). The tensor $\tilde{K}$ solution of 3.9 is now, instead of
5.3,
\begin{equation}
\tilde{K}^{ij}\equiv(\mathcal{\tilde{L}}_{\gamma,conf}X)^{ij}-U^{ij}.
\end{equation}
Noting that if we conformally change the metric via $\gamma^{\prime}%
=\theta^{\frac{4}{n-2}}\gamma$ and the lapse via $\tilde{N}=\theta^{\frac
{-2n}{n-2}}\tilde{N}^{\prime}$ then
\begin{equation}
(\mathcal{\tilde{L}}_{\gamma^{\prime},conf}X)^{ij}=\theta^{-2(n+2)/(n-2)}%
(\mathcal{\tilde{L}}_{\gamma,conf}X)^{ij},
\end{equation}
we find that $K$ has the required scaling.

We are thus led to the following corollary to the theorem 4.2, under otherwise
the same hypothesis.

\begin{corollary}
If the tensor $\tilde{K},$ a solution of the momentum constraint conformally
formulated in a metric $\gamma,$ is obtained as the sum of a given traceless
tensor $U$ and the product by a given function $\tilde{N}$ of a conformal Lie
derivative of a vector $X:$
\begin{equation}
\tilde{K}^{ij}\equiv(\mathcal{\tilde{L}}_{\gamma,conf}X)^{ij}-U^{ij},
\end{equation}
then the tensor
\begin{equation}
\tilde{K}^{\prime ij}\equiv(\mathcal{\tilde{L}}_{\gamma^{\prime},conf}%
X)^{ij}-U^{\prime ij},\text{ \ \ }U^{\prime ij}=\theta^{-2(n+2)/(n-2)}%
U^{ij},\text{ \ }\tilde{N}^{\prime}=\theta^{\frac{2n}{n-2}}\tilde{N}%
\end{equation}
is a solution of the momentum constraint conformally formulated in the metric
$\gamma^{\prime}.$
\end{corollary}

\section{Asymptotically Euclidean Manifolds.}

\subsection{Definitions.}

In the following sections we will study the solution of the conformally
formulated constraints 4.4 and 4.7 on asymptotically euclidean manifolds of
dimension $n\geq3.$

The Euclidean space $E^{n}$ is the manifold $R^{n}$ endowed with the Euclidean
metric, which is $\sum(dx^{i})^{2}$ in canonical coordinates.

A $C^{\infty}$, $n$-dimensional, Riemannian manifolds $(M,e)$ is called
''Euclidean at infinity'' if there exists a compact subset $S$ of $M$ such
that $M-S$ is the disjoint union of a finite number of open sets $U_{i}$, with
each $(U_{i},e)$ being isometric to the exterior of a ball in ${}R^{n}$. Each
open set $U_{i}\subset M$ is sometimes called an ''end'' of $M$. If $M $ is
diffeomorphic to ${}R^{n}$, it has only one end; and we can then take for $e$
the Euclidean metric. Unless otherwise specified our manifolds are without
boundary; hence the manifold $(M,e)$ is complete\footnote{{\footnotesize For
studies on asymptotically Euclidean manifolds with boundary see \ Chrusciel
and Delay [Chru-De], [Ma03], [Ma04a] The articles [Da], [Da-Fr] also consider
such manifolds, using the Friedrich's conformal compactification.}}.

A Riemannian manifold $(M,\gamma)$ is called \textbf{asymptotically Euclidean}
if there exists a Riemannian manifold $(M,e),$ Euclidean at infinity, and if
$\gamma$ tends to $e$ at infinity in each end. Consider one end $U$ and the
canonical coordinates $x^{i}$ in the space ${}R^{n}$ which contains the
exterior of the ball to which $U$ is diffeomorphic. Set $r\equiv\{\sum
(x^{i})^{2}\}^{1/2}$. In the coordinates $x^{i}$ the metric $e $ has
components $e_{ij}=\delta_{ij}$. The metric $\gamma$ tends to $e$ at infinity
if in these coordinates $\gamma_{ij}-\delta_{ij}$ tends to zero. A possible
way of making this statement mathematically precise is to use the Nirenberg -
Walker weighted Sobolev spaces. One can also use in these elliptic constraint
problems weighted H\"{o}lder spaces\footnote{{\footnotesize See [CB-CS].}},
but they are not well adapted to the related evolution questions.

A \textbf{weighted Sobolev space} $W_{s,\delta}^{p}\;,\;$with $1\leq
p\leq\infty,$ with $s$ a positive or zero integer, $\delta$ a real number${}$,
for tensors of some given type on the manifold $(M,e)$ euclidean at infinity
is the space of tensors of that type which admit generalized $e$ - covariant
derivatives of order up to $s$ and for which the following norm is finite:
\begin{equation}
\Vert u\Vert_{W_{s,\delta}^{p}}=\left\{  \sum_{0\leq m\leq s}\int_{V}%
\mid\partial^{m}u\mid^{p}(1+d^{2})^{\frac{1}{2}p(\delta+m)}d\mu\right\}
^{1/p}.
\end{equation}
Here $\partial$, $|\;\;|$ and $d\mu$ denote the covariant derivative, norm and
volume element corresponding to the metric $e$, and $d$ is the distance in the
metric $e$ from a point of $M$ to a fixed point. If $(M,e)$ is a euclidean
space one can choose $d=r$, the euclidean distance to the origin. The space
$\mathcal{D}$ of $C^{\infty}$ tensors with compact support is dense in
$W_{s,\delta}^{p},$ regardless of what $s$ and $\delta$ are, so long as
$p<\infty.$

If $s$ and $\delta$ are large enough, a function (or tensor field) in
$W_{s,\delta}^{p}$ is continuous and tends to zero at infinity. Specifically
if we define $C_{\beta}^{m}$ to be the Banach space of weighted $C^{m}$
functions (or tensor fields) on $(M,e)$ with norm given by
\[
\Vert u\Vert_{C_{\beta}^{m}}\equiv\sum_{0\leq\ell\leq m}\sup_{M}%
(|\partial^{\ell}u|(1+d^{2})^{\frac{1}{2}(\beta+\ell)}),
\]
then the following inequality holds\footnote{{\footnotesize For proofs of this
embedding and the multiplication rule 6.3, see [CB-Ch] 1981, or [CB- DM] II p.
396.}}, with $C$ a number depending only on $(M,e),$
\begin{equation}
\Vert u\Vert_{C_{\beta}^{m}}\leq C||u||_{W_{s,\delta}^{p}},\text{ \ \ if
\ \ }s>m+\frac{n}{p},\text{ \ \ }\delta>\beta-\frac{n}{p}.
\end{equation}
We see that $u\in W_{s,\delta}^{p}$ implies that $u$ is continuous and tends
to zero at infinity if $s>\frac{n}{p}$ and $\delta>-\frac{n}{p}.$

Let $(M,e)$ be a manifold which is Euclidean at infinity. The Riemannian
manifold $(M,\gamma)$ is said to be $(p,\sigma,\rho)$ asymptotically euclidean
if $\gamma-e\in W_{\sigma,\rho}^{p}$. If $\gamma-e\in W_{\sigma,\rho}^{p}$
with $\sigma>\frac{n}{p}$, and $\rho>-\frac{n}{p},$ then $\gamma$ is $C^{0}$
and $\gamma-e$ tends to zero at infinity. The set of Riemannian metrics (i.e.
positive definite symmetric 2-tensors) such that $\gamma-e\in W_{\sigma,\rho
}^{p}$ is denoted $M_{\sigma,\rho}^{p}.$

We recall the multiplication lemma
\begin{equation}
W_{s_{1},\delta_{1}}^{p}\times W_{s_{2,\delta_{2}}}^{p}\subset W_{s,\delta
}^{p}\text{ \ \ if \ \ }s<s_{1}+s_{2}-\frac{n}{p},\text{ \ }\delta<\delta
_{1}+\delta_{2}+\frac{1}{p},
\end{equation}
and the interpolation\footnote{{\footnotesize CB, to appear.}} inequality: for
any $\varepsilon>0,$ there is a $C(\varepsilon)$ such that, for all $u\in
W_{m,\delta}^{p},$ $1\leq p\leq q\leq\infty,$ and $j<m,$ one has
\begin{equation}
||\partial^{j}u||_{_{W_{0,\delta+\frac{n}{p}-\frac{n}{q}+j}^{q}}}\leq
C\{\varepsilon||u||_{W_{m,\delta}^{p}}+C_{\varepsilon}||u||_{W_{0,\delta}^{p}%
}\}.
\end{equation}

\subsection{Linear elliptic systems.}

We state\footnote{{\footnotesize The theorem relies on previous results of
Nirenberg and Walker [N-W], Cantor [Ca], Choquet-Bruhat and Christodoulou
[CB-Ch], and the interpolation lemma to lower the regularity required of
coefficiemts [CB].}} the following existence theorem for solutions of linear ellipticPDE's.

\begin{theorem}
Hypotheses:

Let $(M,e)$ be a smooth Riemannian manifold Euclidean at infinity. Let
\begin{equation}
Lu\equiv a_{2}\partial^{2}u+a_{1}\partial u+a_{0}u
\end{equation}
be a second order linear elliptic operator\ acting on tensor fields on
$(M,e),$ which in terms of components $u^{A},$ $A=1,...p,$ of the tensor $u$
takes the form
\[
(Lu)^{A}\equiv a_{2,B}^{ij,A}\partial_{ij}^{2}u^{B}+a_{1,B}^{i,A}\partial
_{i}u^{B}+a_{0,B}^{A}u^{B}.
\]
Let the principal symbol{\footnotesize \ }$a_{2}^{ij}\xi_{i}\xi_{j}$ of $L$ be
an isomorphism from $R^{p}$ onto $R^{p}$ for $\xi\not =0.$ Suppose the
coefficients of $L$ satisfy the following hypotheses$\;$%
\begin{equation}
a_{2}-A\in W_{2,\delta}^{p}\;,\;a_{1}\in W_{1,\delta+1}^{p},\text{ \ \ }%
a_{0}\in W_{0,\delta+2}^{p}.
\end{equation}
where $A\partial^{2}$ is an elliptic operator with $C^{\infty}$ coefficients,
constant in each end of $(M,e)$ and
\begin{equation}
p>\frac{n}{2},\ \;-\frac{n}{p}<\delta<-\frac{n}{p}+n-2\;.
\end{equation}

Conclusions:

1. The operator $L$ is a continuous mapping from $W_{2,\delta}^{p}$ into
$W_{0,\delta+2}^{p}$

2. a. There exists a number $C_{L}>0,$ depending only on $A$ and on the norms
of $a_{2}-A,$ $a_{1},a_{0}$, and a number $\delta^{\prime}>\delta$ such that
the following inequality holds for all $u\in W_{2,\delta}^{p}:$
\begin{equation}
\Vert u\Vert_{W_{2,\delta}^{p}}\leq C_{L}\{\Vert Lu\Vert_{W_{0,\delta+2}^{p}%
}+||u||_{W_{1,\delta^{\prime}}^{p}}\}\;.
\end{equation}

b. If in addition $L$ is injective there exists \ a number $C$ such that the
following inequality holds for all $u\in W_{2,\delta}^{p}:$
\begin{equation}
\Vert u\Vert_{W_{2,\delta}^{p}}\leq C\Vert Lu\Vert_{W_{0,\delta+2}^{p}}\;.
\end{equation}

c. The operator $L$ has finite dimensional kernel and closed range.

3. a. If the adjoint operator $L^{\ast}$ is injective on $W_{2,\delta}^{p}, $
then $L$ is surjective from $W_{2,\delta}^{p}$ \ onto $W_{0,\delta+2}^{p}.$

b. If $L$ and $L^{\ast}$ are both injective then they are isomorphisms from
$W_{2,\delta}^{p}$ \ onto $W_{0,\delta+2}^{p}.$
\end{theorem}

\begin{corollary}
If, in addition$\;$\ to the previous hypothesis (including injectivity) it
holds that
\begin{equation}
a_{2}-A\in W_{s+2,\delta}^{p}\;,\;a_{1}\in W_{s+1,\delta+1}^{p},\text{
\ \ }a_{0}\in W_{s,\delta+2}^{p}.
\end{equation}
then $L$ is an isomorphism from $W_{s+2,\delta}^{p}$ onto $W_{s,\delta+2}^{p}.$
\end{corollary}

We recall also the following lemma (lemma 5.2 of CB-Ch).

\begin{lemma}
Suppose that $u\in W_{2,\delta}^{p}$ is a solution of the equation
\begin{equation}
Lu\equiv a_{2}\partial^{2}u+a_{1}\partial u+a_{0}u=f
\end{equation}
where $L$ satisfies the hypotheses of the above theorem and where $f\in
W_{0,\tilde{\delta}+2}^{p},$ $\delta\leq\tilde{\delta}.$ Then $u$\ is in fact
in $W_{2,\tilde{\delta}}^{p},$ so long as $\tilde{\delta}<n-2-\frac{n}{p}.$
\end{lemma}

\subsection{The Poisson operator, $\Delta_{\gamma}-a$}

\begin{theorem}
Let $(M,\gamma)$ be a $M_{2,\delta}^{p}$ manifold\footnote{{\footnotesize The
hypothesis p%
$>$%
n/2 is stronger than necessary but simplifies the proof, and is needed later
in our treatment of non linear equations.}}, $p>\frac{n}{2},$ $\delta
>-\frac{n}{p}.$ Let $a\in W_{0,\delta+2}^{p}$be given. The Poisson operator
$\triangle_{\gamma}-a$ is an isomorphism from $W_{2,\delta}^{p}$ onto
$W_{0,\delta+2}^{p}$ if
\begin{equation}
\int_{M}\{|\partial u|^{2}+au^{2}\}\mu_{\gamma}>0,\text{ \ }%
\end{equation}
\ for any \ $u\in W_{2,\tilde{\delta}}^{p},\ $with $\tilde{\delta}$ some
number such that $n-2-\frac{n}{p}>\tilde{\delta}>-1+\frac{n}{2}-\frac{n}%
{p}\;(\tilde{\delta}=-1$ if $p=2),\ $\ with $\ u\not \equiv0.$
\end{theorem}

\begin{proof}
The operator $\triangle_{\gamma}-a$ is self adjoint. It is an isomorphism
$W_{2,\delta}^{p}\rightarrow W_{0,\delta+2}^{p}$ if injective. By lemma 6.3 it
is sufficient to prove the injectivity on $W_{2,\tilde{\delta}}^{p}$ for some
$\tilde{\delta}$ such that $\tilde{\delta}<$ $n-2-\frac{n}{p}.$ The theorem is
obtained by integration on $M$ of $u(\triangle_{\gamma}u-au),$ trivially in
the case $p=2,$ $\tilde{\delta}=-1$ (compatible with $p>\frac{n}{2}$ if and
only if $n=3)$; and by using either Sobolev embeddings or the Holder
inequality in the case $p\not =2$ and $\tilde{\delta}>-1+\frac{n}{2}-\frac
{n}{p}.$
\end{proof}

\begin{theorem}
Let $u$ satisfy the equation
\begin{equation}
\Delta_{\gamma}u-au=-f
\end{equation}
with $\gamma\in M_{2,\delta}^{p},\ \delta>-\frac{n}{p},$ and $p>\frac{n}{2}.$
Suppose $a\in W_{0,\delta+2}^{p},$ $u-c\in W_{2,\tilde{\delta}}^{p}$, where
$c$ is a given number, $-1+\frac{n}{2}-\frac{n}{p}<\tilde{\delta}$ ,
($\tilde{\delta}\geq-1$ if $p=2).$ Suppose$\ \ a\geq0.$ Then $u\geq0$ on $M$
if $f\geq0$ and $\ c\geq0.$ If $f\leq0$ and $\ c\leq0,$ then $u\leq0$ on $M.$
The lower bound of $\tilde{\delta}$ can be weakened to $\delta>-\frac{n}{p}$
if $c=0$ and $f\in W_{2,\tilde{\delta}}^{p},$ $\tilde{\delta}>-\frac{n}%
{p}+\frac{n}{2}-1,$ ($\tilde{\delta}=-1$ if $p=2).$
\end{theorem}

\begin{proof}
The integration on $M$ of $v(\triangle_{\gamma}u-au),$ and the choice
$v=u^{+}=Sup(u,0),$ gives $u^{+}=constant,$ therefore $u^{+}\equiv0$ since
$u^{+}$ tends to zero at infinity.
\end{proof}

\section{Solution of the momentum constraint.}

Given the riemannian metric $\gamma$ and the scalar field $\tilde{N}$ the
conformally formulated momentum constraint reads
\begin{equation}
D_{j}(\mathcal{\tilde{L}}X)^{ij}\equiv(\tilde{\Delta}_{\gamma,conf}%
X)^{i}=F^{i}(\varphi)
\end{equation}
with
\begin{equation}
F^{i}(\varphi)\equiv D_{j}U^{ij}+\frac{n-1}{n}\varphi^{2n/(n-2)}\gamma
^{ij}\partial_{j}\tau+\gamma^{ij}(\partial_{j}\bar{\psi})\tilde{\pi}.
\end{equation}
where $\tau$ is a given function on $M$ and $U$ is a given symmetric traceless
2 - tensor field. The sources $\bar{\psi}$ and $\tilde{\pi}$ are given. We
suppose momentarily that $\varphi$ is also a known function. In fact it
disappears from the equation if $\partial\tau\equiv0$.

\begin{lemma}
Let $(M,\gamma)$ be a $W_{2,\delta}^{p}$ asymptotically euclidean manifold,
and let $\tilde{N}=1+\nu,$ $\nu\in W_{2,\delta}^{p},$ $\tilde{N}>0,$ be given.
Suppose that $p>\frac{n}{2},$ $\delta>-\frac{n}{p};$ then

1. The operator $\tilde{\Delta}_{\gamma,conf}$ is elliptic.

2. Its kernel in W$_{2,\delta}^{p}$ is the space of W$_{2,\delta}^{p}%
$\ conformal Killing vector fields of the metric $\gamma.$
\end{lemma}

\begin{proof}
It holds that
\begin{equation}
(\tilde{\Delta}_{\gamma,conf}Y)^{j}\equiv\tilde{N}^{-1}D_{i}[D^{i}Y^{j}%
+D^{j}Y^{i}-\frac{2}{n}\gamma^{ij}D_{k}Y^{k}]-\tilde{N}^{-2}(\mathcal{L}%
_{\gamma,conf}Y)^{ij}D_{i}\tilde{N}.
\end{equation}
Using the Ricci identity we find that the principal part is
\begin{equation}
\tilde{N}^{-1}[(\Delta_{\gamma}Y)^{j}+(1-\frac{2}{n})D^{j}D_{i}Y^{i}].
\end{equation}
The principal symbol is easily checked to be an isomorphism of $R^{n}$ , for
any $n\geq2.$

2. We prove the second part of this lemma using integration by parts, using
\ lemma 6.3.
\end{proof}

We can now prove the following theorem.

\begin{theorem}
Let $(M,\gamma)$ be a $M_{2,\delta}^{p}$ asymptotically euclidean manifold,
with $p>\frac{n}{2},$ and $\delta>-\frac{n}{p}$ Let $\bar{\psi}\in
W_{2,\delta}^{p}$ and $U$, $\tau,\tilde{\pi}\in W_{1,\delta+1}^{p}$ be given.
Suppose also that $\varphi$\ is known, with $\varphi>0,$ and $(1-\varphi)\in
W_{2,\delta}^{p}$. Then the momentum constraint 4.7 has one and only one
solution $X\in W_{2,\delta}^{p}$ if, in addition, $\delta<n-2-\frac{n}{p}$.

If $\partial\tau\equiv0,$ the condition on $\varphi$ is irrelevant.
\end{theorem}

\begin{corollary}
If in addition $\gamma\in M_{2+s,\delta}^{p},$ $\bar{\psi}\in W_{2+s,\delta
}^{p}$ and $U,\tau,\tilde{\pi}\in W_{1+s,\delta+1}^{p}$ and $(1-\varphi)\in
W_{s+2,\delta}^{p}$ then the solution $X$ belongs to $W_{s+2,\delta}^{p}.$
\end{corollary}

\begin{proof}
The given hypothesis and the Sobolev embedding and multiplication properties
imply that the coefficients of the operator $\tilde{\Delta}_{\gamma,conf}$
satisfy the hypotheses of the theorem 8.2 and that $F(\varphi)\in
W_{0,\delta+2}^{p}.$ The operator $\tilde{\Delta}_{\gamma,conf\text{ }}$ is
self adjoint, and its kernel in $W_{2,\delta}^{p}$ is empty, because there are
no such conformal Killing fields on $(M,\gamma).$
\end{proof}

\section{Solution of the Lichnerowicz equation.}

We consider 4.4\ (the Lichnerowicz equation)
\begin{equation}
\mathcal{H(}x,X,\varphi)\equiv\Delta_{\gamma}\varphi-f(x,\varphi)=0,\text{ \ }%
\end{equation}
where
\[
f(x,\varphi)\equiv r\varphi-a\varphi^{-\frac{3n-2}{n-2}}+b\varphi^{\frac
{n+2}{n-2}},
\]
and
\begin{equation}
r\equiv k_{n}(R(\gamma)-|D\bar{\psi}|_{\gamma}^{2}),\text{ \ \ \ }a\equiv
k_{n}(|\tilde{K}|_{\gamma}^{2}+|\tilde{\pi}|_{\gamma}^{2})\geq0,\text{
\ }b\equiv\frac{n-2}{4n}\tau^{2}-\frac{n-2}{n-1}V(\bar{\psi}).
\end{equation}

We first prove the following lemma.

\begin{lemma}
If $(M,\gamma)$ is a $M_{2,\delta}^{p}$ manifold with $p>\frac{n}{2},$
$\delta>-\frac{n}{p},$ and $\bar{\psi}\in W_{2,\delta}^{p},$ then
$r(\gamma,\bar{\psi})\in W_{0,\delta+2}^{p}.$
\end{lemma}

\begin{proof}
$R(\gamma)$ is a sum of terms of the form $\gamma\partial^{2}\gamma,$ and
$\gamma\partial\gamma\partial\gamma.$ with $\gamma-e\in W_{2,\delta}^{p},$
$\partial\gamma\in W_{1,\delta+1}^{p},$ and $\partial^{2}\gamma\in
W_{\sigma-2,\delta+2}^{p}.$ Under the hypotheses made on $p$ and $\delta,$ the
Sobolev embedding theorem shows that $\gamma-e$ is continuous and bounded on
$M;$ the multiplication theorem completes the proof, also for $|D\bar{\psi
}|_{\gamma}^{2}$.
\end{proof}

\subsection{General existence theorem.}

The following theorem extends to asymptotically Euclidean manifolds a theorem
which has been proved for data on compact\footnote{{\footnotesize Results of
this sort on compact manifolds were proven by Choquet-Bruhat and Leray [CB-Le]
1972, using Leray - Schauder degree techniques. In later work [Is], sub and
super solution techniques have been used.}}manifolds. It can be proved by
similar methods.

\begin{theorem}
Let $(M,\gamma)$ be in $M_{2,\delta}^{p},$ $\delta>-\frac{n}{p},$ $p>\frac
{n}{2}$. Suppose that $a,$ $b,$ $r\in W_{0,\delta+2}^{p},$ and $-1+\frac{n}%
{2}$ $-\frac{n}{p}<\delta$ (if $p=2$ then $\delta=-1$ is admissible). Suppose
the Lichnerowicz equation 4.4 admits a subsolution $\varphi_{-}$ and a
supersolution $\varphi_{+},$ which are continuous and bounded functions with
$\partial\varphi_{+},$ $\partial\varphi_{-}\in W_{1,\delta+1}^{p},$ such that
\begin{equation}
\triangle_{\gamma}\varphi_{-}\geq f(.,\varphi_{-})\;,\quad\triangle_{\gamma
}\varphi_{+}\leq f(.,\varphi_{+}),
\end{equation}%
\begin{equation}
\lim_{\infty}\varphi_{-}\leq1\;,\quad\lim_{\infty}\varphi_{+}\geq1
\end{equation}
and for which there exist numbers $\ell$ and $m,$ with $\ell>0$ if
$a\not \equiv0,$ such that on $M$%
\begin{equation}
\ell\leq\varphi_{-}\leq\varphi_{+}\leq m.\text{\ }%
\end{equation}

Then the equation admits a solution $\varphi$ such that:
\begin{equation}
\varphi_{-}\leq\varphi\leq\varphi_{+}\;,\quad1-\varphi\in W_{2,\delta}^{p}%
\end{equation}
for
\begin{equation}
\delta<n-2-\frac{n}{p}.
\end{equation}

If moreover $\gamma\in M_{2+s,\delta}^{p},$ and $a,$ $b\in W_{s,\delta+2}%
^{p},$ then the solution is such that $1-\varphi\in W_{s+2,\delta}^{p}$.
\end{theorem}

Note that constant sub and super solutions are not natural in the
asymptotically Euclidean case. In our application of this theorem to the
Lichnerowicz equation, we introduce some intermediate steps to obtain non
constant sub and supersolutions.

\subsection{Uniqueness theorem.}

The uniqueness of a solution $\varphi$ of the Lichnerowicz equation follows
from monotonicity if we assume that $r\geq0,a\geq0,$ and $b\geq0.$ A proof of
uniqueness can be given under the same hypothesis on $a$ and $b,$ but with no
restriction on the sign of $r.$

\begin{theorem}
The Lichnerowicz equation 4.4 on $(M,\gamma),$ with $\gamma\in M_{2,\delta
}^{p},$ $p>\frac{n}{2},$ $\delta>-\frac{n}{p}$ has at most one positive
solution $\varphi,$ $\varphi-1\in W_{2,\delta}^{p},$ if $a,b,r\in
W_{0,\delta+2}^{p},$ and if $a\geq0$, and $b\geq0$.
\end{theorem}

\begin{proof}
Suppose it admits two solutions $\varphi_{1}>0$ and $\varphi_{2}>0$. Using the
identity 3.2 we find that, with $\gamma_{i}:=\varphi_{i}^{4/(n-2)}\gamma, $
and $r_{i}:=k_{n}(R(\gamma_{i})-|D\bar{\psi}|_{\gamma_{i}}^{2}),$ $i=1,2$
\begin{equation}
\Delta_{\gamma_{2}}(\varphi_{1}\varphi_{2}^{-1})-(\varphi_{1}\varphi_{2}%
^{-1})r_{2}\equiv-(\varphi_{1}\varphi_{2}^{-1})^{(n+2)/(n-2)}r_{1}%
\end{equation}
\smallskip Since $\varphi_{1}$ is a solution of 8.1 we have
\begin{align*}
r_{1}  &  \equiv-\varphi_{1}^{-(n+2)/(n-2)}\{\Delta_{\gamma}\varphi
_{1}-\varphi_{1}r(\gamma,\bar{\psi})\}\\
&  =\varphi_{1}^{-(n+2)/(n-2)}\{a\varphi_{1}^{(-3n+2)/(n-2)}-\varphi
^{(n+2)/(n-2)}b\}
\end{align*}
\smallskip and an analogous equation for $r_{2}.$ Inserting these results in
the previous equation gives an equation of the form
\begin{equation}
\Delta_{\gamma_{2}}(\varphi_{1}\varphi_{2}^{-1}-1)-\lambda\{(\varphi
_{1}\varphi_{2}^{-1}-1\}=0,
\end{equation}
with
\begin{align}
\lambda &  \equiv a\varphi_{1}^{(-3n+2)/(n-2)}\varphi_{2}^{-(n+2)/(n-2)}%
{\frac{(\varphi_{1}\varphi_{2}^{-1})^{4(n-1)/(n-2)}-1}{\varphi_{1}\varphi
_{2}^{-1}-1}}+\nonumber\\
&  b\varphi_{1}\varphi_{2}^{-1}{\frac{(\varphi_{1}\varphi_{2}^{-1}%
)^{4/(n-2)}-1}{\varphi_{1}\varphi_{2}^{-1}-1}.}%
\end{align}
So long as $\varphi_{1}$ and $\varphi_{2}$ are continuous and positive
functions on $M,$ the fractions with denominator $\varphi_{1}\varphi_{2}%
^{-1}-1$ are continuous and positive functions as well, since the powers of
$\varphi_{1}\varphi_{2}^{-1}$ appearing in their numerators are greater than
1, Therefore $\lambda\in W_{0,\delta+2}^{p}$. Noting that by definition
$a\geq0,$ it follows that if $\tau$ and $V(\bar{\psi})$ are such that
$b\geq0,$ then $\lambda\geq0.$ Hence using $\varphi_{1}\varphi_{2}^{-1}-1\in
W_{2,\delta}^{p}$, and the injectivity of $\Delta_{\gamma}-\lambda$ on
$W_{2,\delta}^{p},$ we have $\varphi_{1}\varphi_{2}^{-1}-1\equiv0.$
\end{proof}

\subsection{Generalized Brill-Cantor Theorem.}

For compact smooth Riemannian manifolds the solutions of the Lichnerowicz
equations have been classified by Isenberg [Is95] through the use of the
Yamabe theorem. The Yamabe conformal invariant is defined by
\[
Inf_{f\in\mathcal{D},f\not \equiv0}(\int_{M}\left\{  |Df|^{2}+k_{n}%
R(\gamma)f^{2}\right\}  \mu_{\gamma}/||f||_{L^{2n/(n-2)}}^{2}\text{.\ \ }%
\]
The Yamabe theorem, proved for smooth metrics in an increasing number of cases
by Trudinger, Aubin and Schoen, says that any compact Riemannian manifold is
conformal to a manifold with constant scalar curvature, $+1,-1,$ or $0$
according to the sign of the Yamabe invariant. It is easy to see that this
theorem extends to $W_{2}^{p}$ metrics, $p>\frac{n}{2},$ in the negative or
zero case. In the positive case only a weaker form (proved by Yamabe himself
in the smooth case) is proved to hold, namely that the $W_{2}^{p}$ manifold is
conformal to a manifold with strictly positive scalar curvature. This property
is used in [CB02] and [Ma04b]. Maxwell in particular establishes the
classification of solutions of the Lichnerowicz equation using only the sign
of the Yamabe invariant and not the full Yamabe theorem. The definition of the
Yamabe conformal invariant extends to non compact manifolds
[Ma03]\footnote{{\footnotesize The definition used by Brill - Cantor in their
theorem, carried over in CB-I-Y, }
\[
\int_{M}\left\{  |Df|^{2}+k_{n}R(\gamma)f^{2}\right\}  \mu_{\gamma
}/||f||_{L^{2n/(n-2)}}^{2}>0\text{{\footnotesize , }}\ \ f\in\mathcal{D}%
,f\not \equiv0\text{\ }%
\]
{\footnotesize was incorrect, because it did not imply this inequality for all
}$f\in W_{2,\delta}^{p},$ {\footnotesize since the limit of positive functions
is not necessarily positive.}} but there is no theorem for asymptotically
euclidean manifolds analogous to the Yamabe theorem, and the denomination of
''positive Yamabe class'' for asymptotically Euclidean manifolds with a
positive Yamabe invariant is somewhat misleading, as shown by the following
theorem, proved\footnote{{\footnotesize Under more restrictive hypothesis on
regularity, and in the case n=3.}} by Brill and Cantor 1981 [Br-Ca] , and
generalized in the presence of a scalar field as follows.

\begin{theorem}
Let $(M,\gamma)$ be a $(p,2,\delta)$ asymptotically Euclidean manifold with
$p>\frac{n}{2},$ $\delta>-\frac{n}{p},$ and let $\bar{\psi}\in W_{2,\delta
}^{p}$ be a scalar field on $M.$ There exists on $M$ a $(p,2,\delta)$
asymptotically euclidean metric $\gamma^{\prime}$ conformal to $\gamma$ such
that $r(\gamma^{\prime},\bar{\psi})=0$ if and only if $(M,\gamma,\bar{\psi})$
satisfy the following inequality\footnote{{\footnotesize This condition,
already used to prove injectivity, is implied by the positivity of the Yamabe
invariant, because $\mathcal{D}$\ is dense in }$W_{2,\delta}^{p}.$}
\begin{equation}
\int_{M}\left\{  |Df|^{2}+r(\gamma,\bar{\psi})f^{2}\right\}  \mu_{\gamma}>0\;
\end{equation}
for every function $f$ on $M$ with $f\in W_{2,\tilde{\delta}}^{p}$,
$\tilde{\delta}>-\frac{n}{p}+\frac{n}{2}-1$ ($\tilde{\delta}\geq-1$ if
$p=2),\ f\not \equiv0.$
\end{theorem}

\begin{proof}
($M,\gamma)$ is conformal to $(M,\gamma^{\prime})\in M_{2,\delta}^{p}$ with
$r(\gamma^{\prime},\bar{\psi})=0$ if and only if there exists a function
$\varphi>0,$ such that $\gamma^{\prime}=\varphi^{4/(n-2)}\gamma\in
M_{2,\delta}^{p}$ and
\begin{equation}
\triangle_{\gamma}\varphi-r(\gamma,\bar{\psi})\varphi=0,
\end{equation}
equivalently, setting $\varphi\equiv1+u,$ 8.12 reads
\begin{equation}
\Delta_{\gamma}u-r(\gamma,\bar{\psi})u=r(\gamma,\bar{\psi}).
\end{equation}

1.\ Suppose that the condition 8.11 is satisfied.The equation 8.13. is linear
and elliptic, with $\Delta_{\gamma}-r(\gamma,\bar{\psi})$ an injective
operator on $W_{2,\delta}^{p}$, and satisfies the hypotheses of the theorem
8.2; therefore it admits a solution $u\in W_{2,\delta}^{p}\subset C_{\alpha
}^{0}.$ It remains to prove that $\varphi\equiv1+u$ is positive, then
$\varphi^{4/(n-2)}\gamma\in M_{2,\delta}^{p}.$ One cannot use directly the
maximum principle because $r(\gamma,\bar{\psi})$ is not necessarily positive.
Inspired by Brill and Cantor (see also [Ma03a]) we consider the family of
equations
\begin{equation}
\triangle_{\gamma}\varphi-kr(\gamma,\bar{\psi})\varphi=0\text{ \ i.e.
\ \ }\triangle_{\gamma}u-kr(\gamma,\bar{\psi})u=kr(\gamma,\bar{\psi})
\end{equation}
with $k\in\lbrack0,1]$ a number. Each of these equations satisfy the condition
8.11 hence admits a solution $u_{k}\in W_{2,\delta}^{p}\subset C_{\alpha}%
^{0},$ and the $C_{\alpha}^{0}$ norm of $u_{k}$ depends continuously on $k.$
The set $S:=\{u_{k}\in C_{\alpha}^{0},$ $u_{k}>-1\}$ is open in $C_{\alpha
}^{0}$ and non empty because for $k=0$ it holds that $u_{0}=0$ (i.e.
$\varphi_{0}=1).$ To show that it is closed, suppose that $u_{k^{\prime}\text{
}}$ belongs to its boundary $\partial S;$ then $u_{k^{\prime}}\geq-1,$
$\varphi_{k^{\prime}}\geq0.$ Suppose that $\varphi_{k^{\prime}},$ solution of
the elliptic equation 8.12, vanishes at a point of $M.$ Then by the weak
Harnack inequality (Trudinger 1973) there is a ball $B_{R}$ of center $x$ and
a number $C$ such that
\begin{equation}
||\varphi_{k^{\prime}}||_{L^{q}(B_{2R})}\leq CInf_{B_{R}}\varphi_{k^{\prime}%
}=0,
\end{equation}
hence $\varphi_{k^{\prime}}=0$ in $B_{R}$ and also, by continuity, on $M.$
This is impossible because $\varphi_{k^{\prime}}$ tends to 1 at infinity.
Hence $\varphi_{k^{\prime}}>0,$ The subset $S$ of $C_{\alpha}^{0}$ being both
open and closed is all of $C_{\alpha}^{0}$.

2.\ Conversely suppose that $\varphi>0$ exists and solves the equation
satisfying the hypothesis of the theorem. Then we will show that for any
$f\not \equiv0$, $f\in W_{2,\delta}^{p}$ the inequality 8.11\ holds. We set
$\theta=f\varphi^{-1},$ then $\theta\in W_{2,\delta}^{p}\subset C_{\alpha}%
^{0}.$ We have by elementary calculus:
\begin{equation}
|Df|^{2}=|D\theta|^{2}\varphi^{2}+\varphi D\varphi.D(\theta^{2})+\theta
^{2}|D\varphi|^{2}\;.
\end{equation}
The following integration by parts holds for the functions under
consideration:
\begin{equation}
\int_{M}\varphi D\varphi.D(\theta^{2})\mu_{\gamma}=\int_{M}-\theta
^{2}D(\varphi D\varphi)\mu_{\gamma}\;.
\end{equation}
Therefore,
\begin{equation}
\int_{M}\varphi D\varphi.D(\theta^{2})\mu_{\gamma}=\int_{M}-\theta^{2}%
(\varphi\triangle_{\gamma}\varphi+|D\varphi|^{2})\mu_{\gamma}%
\end{equation}
and \
\begin{equation}
\int_{M}|Df|^{2}\mu_{\gamma}=\int_{M}\{|D\theta|^{2}\varphi^{2}-\theta
^{2}\varphi\triangle_{\gamma}\varphi\}\mu_{\gamma}\;.
\end{equation}
\ We have $D\theta\not \equiv0$ since $\theta\in C_{\alpha}^{0}$ tends to zero
at infinity and cannot be a constant without being identically zero, which is
ruled out by the hypothesis $f\not \equiv0$. Hence when $\varphi>0$ satisfies
the equation 8.12 the function $f\in W_{2,\delta}^{p},$ $f\not \equiv0$
satisfies the inequality
\begin{equation}
\int_{M}\left\{  |Df|^{2}+r(\gamma,\bar{\psi})f^{2}\right\}  \mu_{\gamma}>0.
\end{equation}
\end{proof}

\textbf{Remark.} The same sort of proof shows that, under the same hypothesis,
there exists on $M$ a metric $\gamma^{\prime}$ conformal to $\gamma$ such that
$r(\gamma^{\prime},\bar{\psi})\leq0.$

\subsection{Existence theorems.}

\begin{theorem}
Let $(M,\gamma)$ be a $M_{2,\delta}^{p}$ manifold with $p>\frac{n}{2}.$ Let
$\bar{\psi}$ be a scalar field on $M$ with potential $V(\bar{\psi}),$ such
that $\bar{\psi}\in W_{2,\delta}^{p}$ and $V(\bar{\psi})\in W_{0,\delta+2}%
^{p}.$ Suppose that 8.11\ is satisfied and $b\geq0.$ The Lichnerowicz
equation
\begin{equation}
\Delta_{\gamma}\varphi-r\varphi+a\varphi^{-\frac{3n-2}{n-2}}-b\varphi
^{\frac{n+2}{n-2}}=0,
\end{equation}
\ \
\begin{equation}
a,b\in W_{0,\delta+2}^{p},\text{ \ }\delta>-1+\frac{n}{2}-\frac{n}{p},\text{
}\delta\geq-1\text{ if }p=2,
\end{equation}
has one and only one solution, $\varphi=1+u,$ $u\in W_{2,\delta}^{p},$ if
$n-2-\frac{n}{p}>\delta>-1+\frac{n}{2}-\frac{n}{p}$ ( extended to $\delta
\geq-1$ if $p=2).$ The solution can be obtained by iteration.
\end{theorem}

\begin{corollary}
If moreover $\gamma\in M_{2+s,\delta}^{p}$ and $a,b\in W_{s,\delta+2}^{p}, $
then $u\in W_{s+2,\delta}^{p}.$
\end{corollary}

\begin{proof}
\textbf{Uniqueness:} This follows from the general theorem 8.3. It can also be
proved directly using the monotonicity of the non linear term.

\textbf{Existence. }Since it follows from theorem 4.2 that the Lichnerowicz
equation is conformally invariant, we may, without loss of generality,
conformally transform equation to a metric such that $r(\gamma,\bar{\psi})=0:
$%
\begin{equation}
\Delta_{\gamma}\varphi+a\varphi^{-\frac{3n-2}{n-2}}-b\varphi^{\frac{n+2}{n-2}%
}=0,
\end{equation}

1. We first consider equation 8.23 with $b=0:$%
\begin{equation}
\Delta_{\gamma}\varphi+a\varphi^{-\frac{3n-2}{n-2}}=0
\end{equation}
This equation admits a constant subsolution $\varphi_{-}=1$ but no finite
constant supersolution. However, it admits a non constant supersolution,
namely the function $\varphi_{+}=1+u_{+}$ with $u_{+}\in W_{2,\delta}^{p}$ a
solution of the linear equation
\begin{equation}
\Delta_{\gamma}u_{+}=-a;
\end{equation}
indeed the maximum principle shows that $u_{+}\geq0,$ hence $\varphi_{+}\geq1$
and
\begin{equation}
\Delta_{\gamma}\varphi_{+}=-a\leq-a\varphi_{+}^{-\frac{3n-2}{n-2}}.
\end{equation}
We can apply the general existence theorem 8.2\ to prove the existence of a
solution $\varphi_{1}.$

2. We next consider the equation with $a=0:$%
\begin{equation}
\Delta_{\gamma}\varphi-b\varphi^{\frac{n+2}{n-2}}=0.
\end{equation}
This equation admits the subsolution $\varphi_{-}=0$ and the supersolution
$\varphi_{+}=1.$ It admits therefore a solution $\varphi_{2},$ with
$1-\varphi_{2}\in W_{2,\delta}^{p},$ and $0\leq\varphi_{2}\leq1.$ We prove
that $\varphi_{2}>0$ by an argument similar to the one used in the proof of
the Brill -Cantor theorem: We consider the family of equations
\begin{equation}
\triangle_{\gamma}\varphi-kb\varphi^{\frac{n+2}{n-2}}=0\text{ }%
\end{equation}
with $k\in\lbrack0,1]$ a number. Each of these equations admits one solution
$\varphi_{k}=1+u_{k}\geq0,$ with $u_{k}\in W_{2,\delta}^{p}\subset C_{\alpha
}^{0},$ and the $C_{\alpha}^{0}$ norm of $u_{k}$ depends continuously on $k.$
The proof continues as in the proof of theorem 8.4.

3. Consider the general equation 8.23. By the above results this admits
$\varphi_{1}$ as a supersolution and $\varphi_{2}$ as a subsolution. Therefore
the existence of a solution follows again from the general existence theorem
8.2. The proof of the corollary also follows from this result.

The proof of the corollary follows from that of theorem 8.2.
\end{proof}

We now state two theorems which suppose $b\leq0.$ They can be applied in
particular when the scalar field has a non negative potential $V(\bar{\psi})$
and the initial manifold is maximal or has an appropriately small mean
extrinsic curvature.

These theorems can also be applied if there exists a density of matter $q$
which is unscaled and non negative. Such a term $q$ adds to $V(\bar{\psi}).$

We first prove a calculus lemma.

\begin{lemma}
Consider the following algebraic function of $y$ with $a>0,r>0$ and $d\geq0: $%
\begin{equation}
f(y)\equiv dy^{\frac{n}{n-2}}-ry^{\frac{n-1}{n-2}}+a
\end{equation}
There are two real numbers $y_{1}$ and $y_{2},$ such that $0<y_{1}\leq y_{2}$
and
\begin{equation}
f(y_{1})\geq0,\text{ \ }f(y_{2})\leq0
\end{equation}
if
\begin{equation}
\text{\ }ad^{n-1}<[\frac{(n-1)^{n-1}}{n^{n}}]r^{n}%
\end{equation}
\end{lemma}

\begin{proof}
Suppose $d>0.$ The function $f$ starts from $a>0$ for $y=0,$ decreases when
$y$ increases from $0$ to $y_{m}=[\frac{(n-1)r}{nd}]^{n-2},$ then increases up
to infinity with $y.$ The numbers $y_{1}$ and $y_{2}$ exist with the indicated
properties if $f(y_{m})<0;$ that is if the inequality 8.31 is satisfied. This
inequality always holds if $d=0:$ $f(y)$ starts then from $a>0$ and decreases
to $-\infty,$ so we can then verify that the numbers $y_{1}$ and $y_{2}$ exist.
\end{proof}

We use this lemma to prove the following result.

\begin{theorem}
Let $(M,\gamma)$ be a $M_{2,\delta}^{p}$ manifold with $p>\frac{n}{2}.$ Let
$\;a,b,r\in W_{0,\delta+2}^{p}$ be given on $(M,\gamma)$, with $a\geq0,$
$r\geq0,$ $b\leq0$ and $\delta>-\frac{n}{p}+\frac{n}{2}-1$ ($\delta\geq-1$ if
$p=2$). The equation
\begin{equation}
\triangle_{\gamma}\varphi-r\varphi+a\varphi^{-\frac{3n-2}{n-2}}-b\varphi
^{\frac{n+2}{n-2}}=0
\end{equation}
has a solution $\varphi>0$, with $1-\varphi\in W_{2,\delta}^{p},$ if
$\delta<n-2-\frac{n}{p}$ and if the inequality 8.31 is satisfied on $M,$ with
$d=-b,$ and so long as
\begin{equation}
\inf_{x\in M}y_{1}(x)>0\;,\quad\inf_{x\in M}y_{2}(x)\geq\max\left\{
1,\sup_{x\in M}z_{1}(x)\right\}  ,
\end{equation}
where $y_{1}(x)$ and $y_{2}(x)$ are the two positive numbers which annul the
algebraic function\footnote{{\footnotesize Polynomial in the case n=3.}}
\begin{equation}
f_{x}(z)\equiv-b(x)y^{\frac{n}{n-2}}-r(x)y^{\frac{n-1}{n-2}}+a(x).
\end{equation}
\end{theorem}

\begin{proof}
The equation admits a constant subsolution $\varphi_{-}=\ell>0$ and a constant
supersolution $\varphi_{+}=m\geq1,$ $\geq\ell,$ and therefore a solution
$\varphi$ with the given properties, so long as almost every $x\in M $ it
holds that
\begin{equation}
f_{x}(\ell^{4})\geq0,\quad f_{x}(m^{4})\leq0.
\end{equation}
The lemma, and the inequalities 8.33 insure the existence of such numbers
$\ell$ and $m,$ given by
\begin{equation}
\ell=\min\left\{  1,\inf_{x\in M}z_{1}(x)\right\}  ,\quad m=\inf_{x\in M}%
z_{2}(x).
\end{equation}
\end{proof}

The next theorem does not rely on the sub - super solution method. It supposes
that $r\leq0,$ hence applies in particular to data satisfying the generalized
positive Yamabe condition, after their conformal transformation to the case
$r=0.$ It has a simpler formulation, but it restricts the size of the
coefficients $a,r$ and $b.$

\begin{theorem}
Let $(M,\gamma)$ be a $M_{2,\delta}^{p}$ manifold with $p>\frac{n}{2}.$ Let
$\;a,b\in W_{0,\delta+2}^{p}$ be given on $(M,\gamma)$, $a\geq0,$ while
$b\leq0,$ $r\leq0,$ $\delta>-\frac{n}{p}+\frac{n}{2}-1$ ($\delta\geq-1$ if
$p=2$). The equation
\begin{equation}
\triangle_{\gamma}\varphi-r\varphi+a\varphi^{-\frac{3n-2}{n-2}}-b\varphi
^{\frac{n+2}{n-2}}=0
\end{equation}
has a solution $\varphi>0$, with $1-\varphi\in W_{2,\delta}^{p},$ if
$\delta<n-2-\frac{n}{p}$ and if $a,b$ and $r$ are small enough in the
$W_{0,\delta+2}^{p}$ norm.
\end{theorem}

\begin{proof}
The equation admits the subsolution $\varphi=1.$ We solve it by iteration,
starting from $u_{0}=1-\varphi_{0}=0.$ We set
\begin{equation}
\triangle_{\gamma}u_{1}=-a+b+r\leq0,
\end{equation}
and we see that $u_{1}$ exists, $u_{1}\geq0,u_{1}\in W_{2,\delta}^{p}$ with
\begin{equation}
||u_{1}||_{W_{2,\delta}^{p}}\leq C_{E}M
\end{equation}
where $C_{E}$ is a number depending only on $\gamma,$ through the constant
$C_{E}$ of the elliptic estimate, and where we have set
\begin{equation}
M:=A+B+R\text{, \ \ \ }A\equiv||a||_{W_{0,\delta+2}^{p}},\text{ \ \ }%
B\equiv||b||_{W_{0,\delta+2}^{p}},\text{ \ \ \ }R\equiv||r||_{W_{0,\delta
+2}^{p}}.
\end{equation}
The Sobolev embedding theorem $W_{2,\delta}^{p}\subset C_{\alpha}^{0}$ implies
then the following inequality where $C_{S}$ is a Sobolev constant
\begin{equation}
||u_{1}||_{C_{\alpha}^{0}}\leq C_{S}||u_{1}||_{W_{2,\delta}^{p}}\leq CM,\text{
\ with \ }C:=C_{S}C_{E}%
\end{equation}
This inequality implies that
\begin{equation}
||\varphi_{1}||_{C^{0}}\leq1+M.
\end{equation}
Recursively, we suppose $u_{n-1}\geq0$ and $||u_{n-1}||_{W_{2,\delta}^{p}}\leq
C_{E}M;$ hence $||u_{n-1}||_{C_{\alpha}^{0}}\leq CM$. The equation defining
$u_{n},$
\begin{equation}
\triangle_{\gamma}u_{n}=-r\varphi_{n-1}+a\varphi_{n-1}^{-\frac{3n-2}{n-2}%
}-b\varphi_{n-1}^{\frac{n+2}{n-2}},
\end{equation}
implies $u_{n}\geq0$ and also that
\begin{equation}
||u_{n}||_{W_{2,\delta}^{p}}\leq C_{E}\{A+R(1+CM)+B(1+CM)^{\frac{n+2}{n-2}}\}.
\end{equation}
Hence $||u_{n}||_{W_{2,\delta}^{p}}\leq C_{E}M\equiv C_{E}(A+B+R)$ if
\begin{equation}
A+R(1+M)+B(1+M)^{\frac{n+2}{n-2}}\leq A+B+R,
\end{equation}
that is:
\begin{equation}
RM+B[(1+M)^{\frac{n+2}{n-2}}-1]\leq0.
\end{equation}
This inequality is satisfied if $A,B,R$ are small enough. The sequence $u_{n}
$ is then uniformly bounded in $W_{2,\delta}^{p}.$ The proof can be completed
by the usual methods of functional analysis.
\end{proof}

\section{Uncoupled system of constraints.}

The conformally formulated momentum and hamiltonian constraints for the
Einstein - scalar field system decouple, in the asymptotically Euclidean case
if the initial manifold $M$ is maximal. When the constraints decouple the
theorems of the previous sections are sufficient to give existence,
non-existence or uniqueness theorems of the systems of constraints. The
previously obtained results give, for example, the following theorems under a
common hypothesis on the a priori given conformal data.

\begin{theorem}
Let $(M,\gamma)$ be a $M_{2,\delta}^{p}$ manifold; $\bar{\psi}\in W_{2,\delta
}^{p}$ a scalar field with potential $V(\bar{\psi})\in W_{0,\delta+2}^{p};$
$\tilde{\pi}\in W_{1,\delta+1}^{p}$ a second scalar field, and $U\in
W_{\delta+1\text{ }}^{p}$ a symmetric 2 - tensor. We assume that $p>\frac
{n}{2},$ $\delta>-1+\frac{n}{2}-\frac{n}{p}$ and $\delta<-2+n-\frac{n}{p}$
($\delta=-1$ is admissible if $p=2).$\textbf{\ }Then the conformally
formulated constraints 7.1 and 8-1 on a maximal submanifold ($\tau=0)$ admit a
solution $X,$ $\varphi=1+u>0,$ with $X,u\in W_{2,\delta}^{p} $ if either the
hypothesis of the theorem 8.5, or 8.8, or 8.9 are satisfied.
\end{theorem}

\begin{proof}
We have already proven that under the given hypotheseses the constraint 7.1
has a unique solution, $X\in W_{2,\delta}^{p},$ therefore $\tilde{K}\in
W_{1,\delta+1}^{p}$ and $a\in W_{0,\delta+2}^{p}$ (Sobolev embedding and
multiplication 6.2, 6.3). We know also (lemma 8.1) that $r\in W_{0,\delta
+2}^{p}.$ Therefore the coefficients of the Lichnerowicz equation (given by
equation 4.5 with $\tau=0)$ satisfy the hypothesis required in the quoted
theorems. It has a solution $\varphi>0,$ $\varphi-1\in W_{2,\delta}^{p}$, and
the pair $X,\varphi$ satisfies the conformally formulated constraints.

This solution is unique in the cases for which the solution of the
Lichnerowicz equation is unique.
\end{proof}

\begin{remark}
The theorem still holds if in addition to the scalar field $\bar{\psi}$ there
exists unscaled sources with zero momentum and energy density $q,$ and we set
$b\equiv-\frac{n-2}{n-1}\{V(\bar{\psi})+q\}.$ This is so because the
constraint equations still decouple (assuming $\tau\equiv0)$ when unscaled
matter sources are present if these sources have a zero
momentum\footnote{{\footnotesize Dain and Nagy [D-N] 2002 consider unscaled
sources with scaled momentum on a maximal submanifold, using H.\ Friedrich
conformal compactification..}}.
\end{remark}

\section{Coupled system of constraints.}

In this section we prove a theorem for the case in which the constraints do
not decouple. This result is in the spirit of a stability theorem.The use of
the implicit function theorem is the simplest way of proving existence of
solutions of equations in the neighbourhood of a given one.

We consider as given the $M_{2,\delta}^{p}$ manifold $(M,\gamma)$ together
with the scalar functions $\bar{\psi},$ $V(\bar{\psi}),$ $\tilde{\pi}$ and the
traceless symmetric 2-tensor $U$, with $\bar{\psi}\in W_{2,\delta}^{p},$
$V(\bar{\psi})\in W_{0,\delta+2}^{p},$ $\tilde{\pi},$ $U\in W_{1,\delta+1}%
^{p}$. We consider the existence of a solution $\varphi$ and $X $ of the
constraints 4.4, 4.7 as we perturb $\tau\in W_{1,\delta+1}^{p}$ away from zero.

We define as follows a mapping $\mathcal{F}$ from open sets of a pair of
Banach spaces into another Banach space:
\begin{equation}
\mathcal{F}\text{: (}W_{1,\delta+1}^{p};W_{2,\delta}^{p}\times W_{2,\delta
}^{p}\cap\varphi>0)\rightarrow W_{0,\delta+2}^{p}\times W_{0,\delta+2}%
^{p},\text{ \ \ }p>\frac{n}{2},\text{ \ }\delta>-\frac{n}{p}%
\end{equation}
by
\begin{equation}
(\tau;X,u\equiv\varphi-1)\mapsto(\mathcal{H}(\tau;\varphi,X),\;\mathcal{M}%
(\tau;\varphi,X))
\end{equation}
where $\mathcal{H}$ and $\mathcal{M}$ are the left hand sides of the conformal
formulation 4.4, 4.7 of the constraints.

The multiplication properties of weighted Sobolev spaces show that
$\mathcal{F}$ is a $C^{1}$ mapping. The partial derivative $\mathcal{F}%
_{X,u}^{\prime}$ at a point $(0;X,u)$ is the linear mapping from $W_{2,\delta
}^{p}\times W_{2,\delta}^{p}$ into $W_{0,\delta+2}^{p}$ given by
\begin{equation}
(\delta X,\delta u)\mapsto(\delta\mathcal{H},\delta\mathcal{M)}%
\end{equation}
where ($\delta b=0$ because $\tau=0$ at the considered point and $V(\bar{\psi
})$ is fixed)
\begin{equation}
\delta\mathcal{H}\equiv\Delta_{\gamma}\delta u-\alpha\delta u+\varphi
^{-\frac{3n-2}{n-2}}\delta a,\text{\ }%
\end{equation}
\ with
\begin{equation}
\alpha=r+\frac{3n-2}{n-2}a\varphi^{-4\frac{n-1}{n-2}}+\frac{n+2}{n-2}%
b\varphi^{\frac{4}{n-2}},
\end{equation}
and, using the expressions for $a$ and $\tilde{K},$%
\begin{equation}
\delta a=\frac{n-2}{2(n-1}\tilde{K}(\mathcal{\tilde{L}}_{\gamma,conf})\delta
X.
\end{equation}
On the other hand
\begin{equation}
\delta\mathcal{M}\equiv(\tilde{\Delta}_{\gamma,conf}\delta X)^{i}.
\end{equation}

\begin{theorem}
Specify on the $M_{2,\delta}^{p}$ manifold $(M,\gamma)$ the scalar functions
$\bar{\psi},$ $V(\bar{\psi}),$ $\tilde{\pi},$ $\tilde{N}$ and the traceless
symmetric 2-tensor $U$, with $\bar{\psi}\in W_{2,\delta}^{p},$ $V(\bar{\psi
})\in W_{0,\delta+2}^{p},$ $\tilde{\pi},$ $U\in W_{1,\delta+1}^{p},$
$\tilde{N}-1\in W_{s+2,\delta}^{p}$, $\tilde{N}>0,\;p>\frac{n}{2}, $
$\ -\frac{n}{p}<\delta<n-2-\frac{n}{p}$. Let $(X_{0},\varphi_{0})$ be a
solution of the corresponding constraints with $\tau_{0}=0$. Suppose that for
some $\tilde{\delta}>-1+\frac{n}{2}-\frac{n}{p}$ ($\tilde{\delta}=-1$ if
$p=2)$ it holds that:
\begin{equation}
\int_{M}\{|Df|_{\gamma}^{2}+\alpha_{0}f^{2}\}d\mu_{\gamma}>0\text{ \ \ for all
\ }f\in W_{2,\tilde{\delta}}^{p},\text{ \ }f\not \equiv0
\end{equation}
with
\begin{equation}
\alpha_{0}:=r+\frac{3n-2}{n-2}a_{0}\varphi_{0}^{-4\frac{n-1}{n-2}}-\frac
{n+2}{n-2}V(\bar{\psi})\varphi_{0}^{\frac{4}{n-2}}\geq0\;.
\end{equation}
Then there exists a neighbourhood $\Omega$ of zero in $W_{1,\delta+1}^{p}$
such that, if $\tau\in\Omega$, the coupled constraints have one and only one
solution $(X,\varphi),\;$with $\ \varphi>0,$ and$\;X,u\equiv\varphi-1\in
W_{2,\delta}^{p}.$
\end{theorem}

\begin{proof}
Under the hypotheses that we have made, due to properties of elliptic
equations discussed above, the partial derivative of $\mathcal{F}$ with
respect to the pair $(u,X)$ determines an isomorphism from $W_{2,\delta}%
^{p}\times W_{2,\delta}^{p}$ onto $W_{0,\delta+2}^{p},$ given by
\begin{equation}
(\delta u,\delta X)\mapsto(\delta\mathcal{M},\delta\mathcal{H)}.
\end{equation}
A straightforward application of the implicit function theorem then completes
the proof.
\end{proof}

\begin{remark}
The conclusion of the theorem holds in particular if $(M,\gamma,\bar{\psi})$
satisfy the inequality 8.11 and $V(\bar{\psi})\leq0,$ since then one has
always $a_{0}\geq0.$
\end{remark}

\begin{remark}
An analogous method can be used to prove the existence of solutions of the
coupled system in the additional presence of matter sources with momentum
small enough in $W_{0,\delta+2}^{p}$ norm.
\end{remark}

\bigskip

\bigskip

\noindent
\textbf{Acknowledgments.} This work was partially done in the Department of
Mathematics of the University of Washington, whose hospitality is gratefully
aknowledged by YCB and JI.

\bigskip
\noindent
This research was partially supported by NSF Grant PHY-0354659 at the
University of Oregon, and by NSF Grant DMS-0305048 at the 
University of Washington.

\bigskip

\bigskip

\noindent YCB: ycb@ccr.jussieu.fr\hfill\break
Acad\'{e}mie des Sciences, 23 Quai Conti, 75270 Paris 06, France

\bigskip

\noindent JI: jim@newton.uoregon.edu\hfill\break
Department of Mathematics, University of Oregon, Eugene, OR.
97403 USA

\bigskip

\noindent DP: pollack@math.washington.edu\hfill\break
Department of Mathematics, University of Washington, Seattle,
WA, 98195-4350, USA

\vfill\eject
\noindent
\textbf{Bibliography}

\bigskip

\noindent
[Au] Aubin, A. (1982)
``Nonlinear analysis on manifolds, Monge Ampere equations.'' Springer-Verlag

\bigskip
\noindent
[Ba] Bartnik, R. (1986) {\em The mass of an asymptotically flat manifold}, Comm. Pure
Appl. Math. {\bf 39} 661--693.

\bigskip
\noindent
[Ba-Is] Bartnik, R.and Isenberg, J.\ (2004) {\em The constraint equations},
\  ``The Einstein equations and the large
scale behaviour of gravitational fields, 50 years of the Cauchy problem in
General Relativity'', P.T. Chrusciel and H. Friedrich ed..

\bigskip
\noindent
[Br-Ca] Brill, D. and  Cantor, M. \ (1981) {\em The Laplacian on asymptotically flat
manifolds and the specification of scalar curvature}, 
Comp. Math. {\bf 43}, 317--324.

\bigskip
\noindent
[CB72] Choquet-Bruhat, Y. (1972) {\em Solutions globales du probl\`{e}me des
contraintes sur une vari\'{e}t\'{e} compacte}, C.R. Acad. Sci. {\bf 274}, 
682--684; and 
{\em Global solutions of the problem of constraints on closed manifolds}, Acta Ist.
alta mat. XII, 317--325.

\bigskip
\noindent
[CB75] Choquet-Bruhat, Y. (1975) {\em The problem of constraints in General 
Relativity}, ``Differential Geometry and Relativity'', volume in honor of 
Lichnerowicz, M. Cahen and M Flato ed., Reidel.

\bigskip
\noindent
[CB93] Choquet-Bruhat, Y. (1993) {\em The coupled Einstein constraints}, 
''Directions in General Relativity'', B.L. Hu and T.\ Jacobson ed. Cambridge
University Press.

\bigskip
\noindent
[CB96] Choquet-Bruhat, Y. (1996) {\em Global existence theorems for 
Einstein equations 
in high dimensions},   ``Gravity, particles and space-time'',  19--28, 
World Sci. Publishing, River Edge, NJ. Sardanishvily ed.

\bigskip
\noindent
[CB02] Choquet-Bruhat, Y. (2004) {\em Einstein constraints on $n$ dimensional
compact manifolds}, Conference at the symposium ``The Einstein equations and
the large scale behaviour of gravitational fields, 50 years of the Cauchy
problem in General Relativity.'' Cargese 2002, published in CQG {\bf 21}, 
127--153.

\bigskip
\noindent
[CB-DM] Choquet-Bruhat, Y. and  DeWitt-Morette, C. 
``Analysis, Manifolds and Physics''
I (1982) and II (2000), North Holland.

\bigskip
\noindent
[CB-CS] Choquet-Bruhat, Y. and Chaljub Simon, A. (1978) {\em Solutions
asymptotiquement euclidiennes de l'\'{e}quation de Lichnerowicz}, 
C. R. Acad. Sci. Paris {\bf 286}, 917--920 
(see also Ann. Univ. Toulouse 1979, {\bf 1}, 25-39).

\bigskip
\noindent
[CB-Is-Mo] Choquet-Bruhat, Y., Isenberg, J. and Moncrief, V. (1992) 
{\em Solutions of
constraints for Einstein equations}, C.R. Acad. Sci. {\bf 315}, 349--355.

\bigskip
\noindent
[CB-Is-Yo] Choquet-Bruhat, Y., Isenberg, J., and York, J.W. (2000) 
{\em Einstein constraints on asymptotically Euclidean manifolds}, Phys Rev D (3) {\bf 61} 
 no. 8, 084034.

\bigskip
\noindent
[CB-Ch] Choquet-Bruhat, Y. and Christodoulou, D. (1981) {\em Elliptic systems in
$H_{s,\delta}$ spaces on manifolds which are Euclidean at infinity}, Acta
Math. {\bf 146} 129--150.

\bigskip
\noindent
[CB-Le] Choquet-Bruhat, Y. and Leray, J. (1972) 
{\em Sur le probl\`{e}me de Dirichlet
quasi lin\'{e}aire d'ordre 2}, C.R. Acad. Sci. {\bf 274}, 81--85.

\bigskip
\noindent
[CB-Yo] Choquet-Bruhat, Y. and York, J.W. (1980) {\em The Cauchy problem}, 
``General
Relativity and Gravitation'', Vol. 1,  pp. 99--172, Plenum, New York-London,
Held ed. 

\bigskip
\noindent
[Ch-OM] Christodoulou, D. and O'Murchada, N. (1980) 
{\em The boost problem in General
Relativity}, Comm. Math. Phys. {\bf 80} 271--300.

\bigskip
\noindent
[Chru-De] Chrusciel, P.T. and Delay, E.\ (2003) 
{\em On mapping properties of the
general relativistic constraints operator in weighted function spaces, with
applications}, arxiv.org/gr-qc/0301073, M\'{e}moires de la Soci\'{e}t\'{e}
Math\'{e}matiques de France, 2004.

\bigskip
\noindent
[Da] Dain, S (2004) 
{\em Trapped surfaces as boundaries for the constraint equations.}
Class. Quant. Grav. {\bf 21} 555--574.

\bigskip
\noindent
[Da-Fr] Dain, S. and Friedrich, H. (2001) {\em Asymptotically flat initial data with
prescribed regularity at infinity}, arxiv.org/gr-qc/0102047,
Comm. Math. Phys.  {\bf 222}  (2001),  no. 3, 569--609. 

\bigskip
\noindent
[Da-Na] Dain, S. and Nagy, G. (2002) {\em Initial data for fluid bodies in General
Relativity}, Phys Rev D {\bf 65} 084020-1-15.

\bigskip
\noindent
[Is87] Isenberg, J. (1987) {\em Parametrization of the space of solutions of
Einstein equarions}, Phys. Rev. Let. {\bf 59}, 2389-2392.

\bigskip
\noindent
[Is95] Isenberg, J. (1995) {\em Constant mean curvature solutions of Einstein
constraint equations on closed manifolds}, Class. Quant. Grav. {\bf 12}, 
2249-2279.

\bigskip
\noindent
[Is-Mo] Isenberg, J. and Moncrief, V.\ (1996) {\em A set of nonconstant mean
curvature solutions of Einstein constraint equations on closed manifolds},
Class. Quant. Grav. {\bf 13}, 1819-1847.

\bigskip
\noindent
[Li] Lichnerowicz, A.\ (1944) 
{\em L'int\'{e}gration des \'{e}quations relativistes
et le probl\`{e}me des n corps}, J.\ Math. pures et app. {\bf 23}, 37-63.

\bigskip
\noindent
[Ma03] Maxwell, D. (2004) {\em Solutions of the Einstein constraint equations with
apparent horizon boundaries}, Comm. Math. Phys. {\bf 253}  (2005),  no. 3, 
561--583. 

\bigskip
\noindent
[Ma04a] Maxwell, D. (2004) {\em Rough solutions of the Einstein constraint equations},
arxiv.org/gr-qc/0405088, to appear J.\ Reine Angew. Math.

\bigskip
\noindent
[Ma04b] Maxwell, D. (2004) {\em Rough solutions of the Einstein constraint equations
on compact manifolds}, 
arxiv.org/gr-qc/0506085, to appear J. Hyper. Diff. Eq. {\bf 2}, no. 2, 2005.

\bigskip
\noindent
[OM-Yo] O'Murchada, N. and York, J.W. (1973) {\em Existence and uniqueness of
solutions of the Hamiltonian constraint of General Relativity on compact
manifolds} J. Math. Phys. {\bf 14}, 1551--1557.

\bigskip
\noindent
[Ni-Wa] Nirenberg, L. and Walker, H. F. (1973) {\em The null spaces of partial
differential operators on $R^{n}$}, J. Math. Anal. App. {\bf 42}, 271-301

\bigskip
\noindent
[Yo72] York ,J.W. (1972) {\em Role of conformal three-geometry in the dynamics of
gravitation}, Phys Rev Lett {\bf 28}, 1082--1085.

\bigskip
\noindent
[Yo99] York, J.W. (1999) {\em Conformal thin sandwich initial data for the initial
value problem of general relativity}, Phys Rev Lett. {\bf 82}, 1350--1353.

\end{document}